\documentclass{article}
\usepackage[left=2cm, right=2cm, top=2cm]{geometry}
\usepackage{amssymb,amsmath,verbatim,mathtools,needspace,enumitem,etoolbox,graphicx,physics,microtype,afterpage,xspace,tabularx,lmodern,multirow}
\usepackage{authblk}
\usepackage{gensymb}
\usepackage[utf8]{inputenc}
\usepackage{tipa}
\usepackage[usenames,dvipsnames]{xcolor}
\definecolor{linkcolor}{rgb}{0.0,0.3,0.5}
\definecolor{romared}{RGB}{142,0,28}
\usepackage[square,sort,comma,numbers]{natbib}
\usepackage[unicode, colorlinks=true, linkcolor=linkcolor, citecolor=linkcolor, filecolor=linkcolor, urlcolor=linkcolor, linktocpage, breaklinks]{hyperref}
\usepackage[all]{hypcap}
\usepackage[T1]{fontenc}
\hypersetup{colorlinks=true,citecolor=romared,linkcolor=linkcolor,urlcolor=linkcolor}

\newcommand{\qmul}{Geometry, Analysis and Gravitation, School of Mathematical Sciences, Queen Mary University of London,
Mile End Road, London E1 4NS, United Kingdom}
\newcommand{\tuebingen}{Theoretical Astrophysics, Eberhard Karls University of T\"ubingen, T\"ubingen 72076, Germany}
\newcommand{\cambridge}{Department of Applied Mathematics and Theoretical Physics, Centre for Mathematical
Sciences, University of Cambridge, Wilberforce Road, Cambridge CB3 0WA, United Kingdom}

\title{GRFolres: A code for modified gravity simulations in strong gravity}
\author[1]{Llibert Arest\'e Sal\'o}
\author[1]{Sam E. Brady}
\author[1]{Katy Clough}
\author[2]{Daniela Doneva}
\author[3]{Tamara Evstafyeva}
\author[1]{Pau Figueras}
\author[1]{Tiago França}
\author[1]{Lorenzo Rossi}
\author[1]{Shunhui Yao}

\affil[1]{\qmul}
\affil[2]{\tuebingen}
\affil[3]{\cambridge}

\begin{document}

\maketitle

The following brief overview has been prepared as part of the submission of the code to the Journal of Open Source Software. The code itself can be found at \url{https://github.com/GRChombo/GRFolres}
\footnote{Folres (pronounced \textit{fol-res}) is a word meaning covers or linings in the Catalan language. It has a specific application in the tradition of \textit{Castells} (Human Towers), denoting the second layers of reinforcement above the base \textit{pinya}. We use it here in analogy to our understanding of effective field theories (EFTs) of gravity as an infinite sum of terms organised as a derivative expansion, in which the first one corresponds to GR (with up to 2 derivatives), and the second one to modified theories up to 4 derivatives, which are those that we are able to simulate with GRFolres.}.

\section{Summary}

Gravitational waves (GWs) are generated by the mergers of dense, compact objects like black holes (BHs) and neutron stars (NSs). This provides an opportunity to study the strong field, highly dynamical regime of Einstein's theory of general relativity (GR) at higher curvature scales than previous observations \cite{LISA:2022kgy,Perkins:2020tra,Barausse:2020rsu, Gnocchi:2019jzp,Barack:2018yly,Baker:2014zba}. It is possible that at such scales modifications to GR may start manifest. However, in order to detect such modifications, we need to understand what deviations could look like in theories beyond GR, in particular in the merger section of the signal in near equal mass binaries, which are key targets of the LIGO-Virgo-KAGRA network of detectors (and their future 3G successors). 
Such predictions necessitate the use of numerical relativity (NR), in which the (modified) equations of GR are evolved from an initial configuration several orbits before merger, through the merger period and the subsequent ``ringdown'', during which the gravitational wave signal can be extracted near the computational boundary. 

Current waveforms are tested for consistency with GR by measuring parameterised deviations to the merger, inspiral and ringdown phases \cite{Maggio:2022hre,Krishnendu:2021fga, LIGOScientific:2021sio,Carson:2019kkh,Cornish:2011ys}, and not by comparison to any particular theories. If we obtain predictions for specific models, we can check whether such parameterised deviations are well-motivated and consistent in alternative theories of gravity \cite{LISA:2022kgy, Okounkova:2022grv, Johnson-McDaniel:2021yge,Shiralilou:2021mfl,Perkins:2021mhb,Carson:2020ter,Carson:2020cqb}, and quantify our ability to constrain model parameters using GW observations. 

There are many ways to modify GR, one of the simplest being to couple an additional scalar degree of freedom to gravity, which may (if certain conditions are satisfied) result in so-called ``hairy'' stationary black hole solutions; that is, black holes with a stable, non trivial configuration of the scalar field around them (see \cite{Doneva:2022ewd} for a review). An example of this is the class of Horndeski models \cite{Horndeski:1974wa}. Cubic Horndeski theories have been studied in \cite{Figueras:2020dzx,Figueras:2021abd} and an implementation of this is included in GRFolres. Another more general example within the Horndeski models is the four-derivative scalar-tensor theory (4$\partial$ST), which is the most general theory with up to fourth powers of the derivatives (but still second order equations of motion). Despite their relative simplicity, many models have lacked well-posed (and thus numerically stable) formulations until recently. 

An important breakthrough was made in 2020 by Kov\'acs and Reall, who showed that Horndeski theories are indeed well-posed in a modified version of the harmonic gauge \cite{Kovacs:2020pns,Kovacs:2020ywu} -- a particular coordinate system already used in NR. Subsequently, several specific theories within these classes were probed in their highly dynamical and fully non-linear regimes \cite{East:2020hgw,East:2021bqk,East:2022rqi,Corman:2022xqg}. The extension of the results of \cite{Kovacs:2020pns,Kovacs:2020ywu} to the alternative ``singularity avoiding'' coordinates in \cite{AresteSalo:2022hua,AresteSalo:2023mmd,Doneva:2023oww} offers an alternative gauge in which to probe questions of hyperbolicity, and may even offer stability advantages for certain cases such as unequal mass ratios, as studied in \cite{Corman:2022xqg}. Numerical work on these theories is still in the early stages of development and many technical details on their numerical implementation need to be further investigated. Equally, many scientific questions, concerning our accurate understanding of binary black holes' phenomenology in alternative theories of gravity and their implications for tests of GR, also remain unanswered.

The goal of GRFolres is to meet this need for further research, and to provide a model code to help others develop and test their own implementations. The code is based on the publicly available NR code GRChombo \cite{Clough:2015sqa,Andrade:2020dgc}, which itself uses the open source Chombo framework \cite{Adams:2015kgr} for solving partial differential equations (PDEs). 

In the following sections, we discuss the key features, motivations, and applications of the code.

\section{Key features}
GRFolres inherits many of the features of GRChombo and Chombo. Here we list the key features.

\begin{itemize}
    \item Stable gauge evolution - The code implements the modified moving puncture gauge that ensures a well-posed evolution in the weak coupling regime, as proposed in \cite{AresteSalo:2022hua}. The precise form of the gauge and its parameters can be changed and the standard moving puncture gauge is safely recovered by setting certain parameters to zero.
    \item Modified gravity theories - The currently available theories in the code are 4$\partial$ST and cubic Horndeski. The code is templated over the theory (in the same way that GRChombo is templated over a matter class) so that it can easily be changed without major code modifications. The code also provides an implementation of 4$\partial$ST without backreaction onto the metric (but including the possibility of using the new gauge), to enable comparison with previous works using the decoupling limit approximation.
    \item Accuracy -- The fields are evolved with a 4th order Runge-Kutta time integration and their derivatives calculated with the same finite difference stencils used in GRChombo (4th and 6th order are currently available).
    \item Boundary Conditions -- GRFolres inherits all the available boundary conditions in GRChombo, namely, extrapolating, Sommerfeld (radiative), reflective, and periodic. 
    \item Initial Conditions -- The current examples use solutions that approximately or trivially solve the modified energy and momentum constraints of the theory. An elliptic solver for more general configurations is under development, using a modified CTTK formalism \cite{Aurrekoetxea:2022mpw, Brady:2023dgu}.
    \item Diagnostics -- GRFolres has routines for monitoring the constraint violation and calculating the energy densities associated with the different scalar terms in the action, as discussed in \cite{AresteSalo:2022hua, AresteSalo:2023mmd, Doneva:2023oww}. Other diagnostics can be added as required. We also extract data for the tensor and scalar gravitational waveforms.
    \item C++ class structure -- Following the structure of GRChombo, the GRFolres code is also written in C++ and uses object oriented programming (OOP) and templating.
    \item Parallelism -- GRChombo uses hybrid OpenMP/MPI parallelism with explicit vectorisation of the evolution equations via intrinsics, and is AVX-512 compliant.
    \item Adaptive Mesh Refinement -- The code inherits the flexible AMR grid structure of Chombo, which provides Berger-Oliger style \cite{Berger:1984zza} AMR with block-structured Berger-Rigoutsos grid generation \cite{Berger:1991}. Depending on the problem, the user may specify the refinement to be triggered by the additional degrees of freedom, i.e. the scalar field, or those of the metric tensor.
\end{itemize}

\section{Statement of Need}
As far as we are aware there is currently no other publicly available code that implements the 4$\partial$ST theory of modified gravity or the cubic Horndeski theory in (3+1)-dimensional numerical relativity. 

There is at least one private code, based on the PAMR/AMRD and HAD \cite{East:2011aa,Neilsen:2007ua} infrastructure, that was used in the first works to successfully implement the modified general harmonic gauge for 4$\partial$ST \cite{East:2020hgw,East:2021bqk,East:2022rqi,Corman:2022xqg}. 
Since this code uses a Generalised Harmonic Coordinates (GHC) formulation, it necessitates excision of the interior of black holes, which can be difficult to implement in practice. As a consequence, many groups in the numerical relativity community have opted to use singularity avoiding coordinates such as the BSSN \cite{Nakamura:1987zz,Shibata:1995we,Baumgarte:1998te}, Z4C \cite{Bona:2003fj,Bernuzzi:2009ex} or CCZ4 \cite{Alic:2011gg,Alic:2013xsa} formulations in the puncture gauge \cite{Campanelli:2005dd,Baker:2005vv}, which do not require the excision of the interior of black holes from the computational domain. 
In GRFolres, we use the results of \cite{AresteSalo:2022hua,AresteSalo:2023mmd,Doneva:2023oww} to extend the well-posed formulations of modified gravity to singularity avoiding coordinates. This provides an alternative gauge to the modified GHC one used by other groups. Not only does this provide a valuable comparison to their work, but also eliminates the need for excision.

There are also a number of (3+1)-dimensional codes that implement the equations for the additional scalar degree of freedom in Einstein-scalar-Gauss-Bonnet without backreaction onto the metric tensor, including one implementation using GRChombo \cite{Evstafyeva:2022rve}, which we have integrated into GRFolres to enable comparison between the methods. In particular, Canuda (\url{https://bitbucket.org/canuda}) \cite{Witek:2018dmd} which uses the Einstein Toolkit (\url{http://einsteintoolkit.org/}), with its related Cactus (\url{http://cactuscode.org}) \cite{Loffler:2011ay,Schnetter:2003rb} and Kranc (\url{http://kranccode.org}) \cite{Husa:2004ip} infrastructure, was used in \cite{Richards:2023xsr,Elley:2022ept,R:2022tqa,Silva:2020omi,Witek:2018dmd}. Another implementation is based on the Spectral Einstein Code or SpEC (\url{http://www. black-holes.org/SpEC.html}) \cite{Pfeiffer:2002wt}, as used in \cite{Okounkova:2020rqw}. A neutron star background was considered in \cite{Kuan:2023trn} with a modification of SACRA-MPI code \cite{Yamamoto:2008js,Kiuchi:2017pte}.
Whilst order-reduced methods like those in \cite{Richards:2023xsr,R:2022tqa,Okounkova:2022grv,Elley:2022ept,Doneva:2022byd,Okounkova:2020rqw,Silva:2020omi,Okounkova:2019zjf,Okounkova:2019dfo,Witek:2018dmd,Evstafyeva:2022rve,Kuan:2023trn} 
provide an estimate of the scalar dynamics and associated energy losses, they may miss information about the fully non-linear impact on the metric and suffer from the accumulation of secular errors over long inspirals.

In spherical symmetry several codes have been developed that implement Einstein-scalar-Gauss-Bonnet (a subset of the $4\partial$ST theory that we include as an example in GRFolres). In particular, using the NRPy framework (\url{http://astro.phys.wvu.edu/bhathome}) \cite{Ruchlin:2017com} in \cite{Doneva:2022byd}, the private code of Ripley \& Pretorius in \cite{R:2022hlf,Ripley:2020vpk,Ripley:2019irj,Ripley:2019hxt,Ripley:2019aqj}, and a modification of the GR1D code \cite{OConnor:2009iuz,Gerosa:2016fri} for calculating core collapse in Einstein-scalar-Gauss-Bonnet in \cite{Kuan:2021lol}. There is also the fully nonlinear spherical code developed in \cite{Corelli:2022pio,Corelli:2022phw}. Spherical codes provide a useful testing ground in which coordinate ambiguities can be avoided \cite{R:2022hlf}, and a well posed formulation is easier to obtain, but they lack the generality required to study objects with angular momentum, or binary mergers.

\section{Research projects to date using GRFolres}
So far the code has been used to study a range of fundamental physics problems, as listed here.
\begin{itemize}

    \item The test field case was used in \cite{Evstafyeva:2022rve} to model the scalar waves produced during the ringdown stage of binary black hole coalescence in Einstein-scalar-Gauss-Bonnet, and quantify the extent to which current and future gravitational wave detectors could observe the spectrum of scalar radiation emitted.

    \begin{figure}[h!]
        \begin{center}
	    \includegraphics[width=0.75\textwidth]{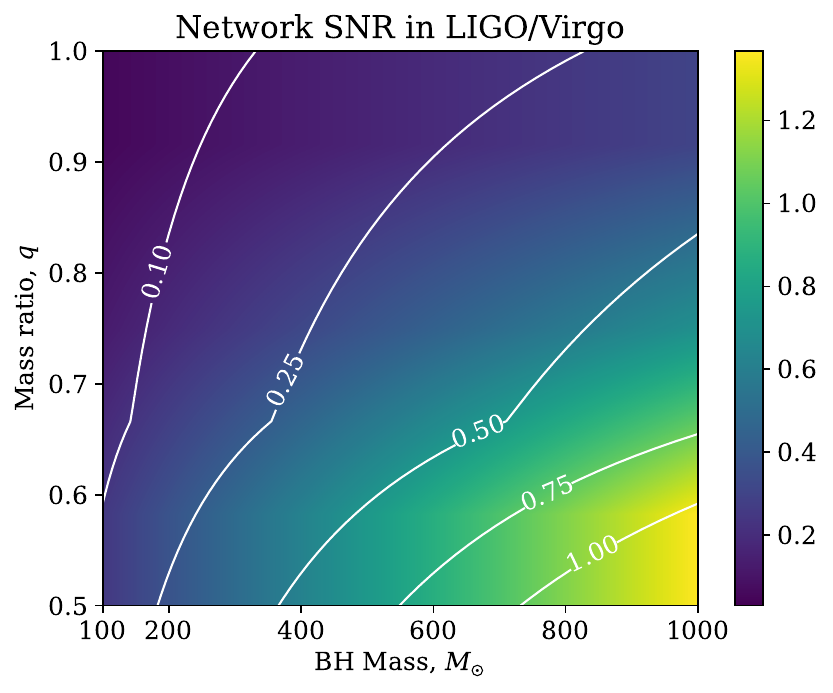}
	    \caption{Contour plot of network signal-to-noise ratio (SNR) for the scalar ringdown of a binary black hole (BBH) in Einstein-scalar-Gauss-Bonnet gravity at 1 Gpc as observed by the Virgo, Livingston and Hanford network of detectors at design sensitivity. Taken from \cite{Evstafyeva:2022rve}.
        }
        \label{fig:testfield}
        \end{center}
    \end{figure}  

    \item The regime of validity of effective field theory in collapse and binary evolutions in cubic Horndeski theories was studied in \cite{Figueras:2020dzx,Figueras:2021abd}. It was found that the mismatch of the gravitational wave strain can be as large as 10\%–13\% in the Advanced LIGO mass range for such theories.

    \begin{figure}[h!]
        \begin{center}
	    \includegraphics[width=0.9\textwidth]{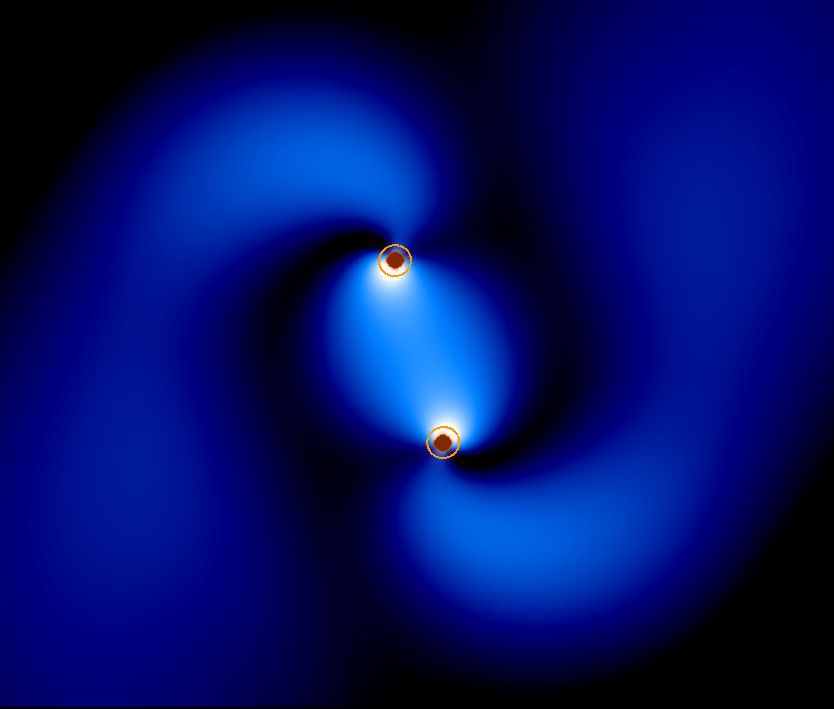}
	    \caption{Energy density (in blue) of the scalar field surrounding the binary black holes for the Horndeski theory at a representative instant of time during the inspiral phase. The apparent horizon of the black holes is shown in orange. The region where the weak coupling conditions are larger than one is depicted in brown. Taken from \cite{Figueras:2021abd}.}
        \label{fig:cubic}
        \end{center}
    \end{figure}  

    \item In the work \cite{AresteSalo:2022hua}, the code was developed and tested, with waveforms for shift-symmetric theories of Einstein-scalar-Gauss-Bonnet gravity produced for equal mass binaries, as illustrated in Fig. \ref{fig:waveform}.

    \begin{figure}[h!]
	    \includegraphics[width=0.99\textwidth]{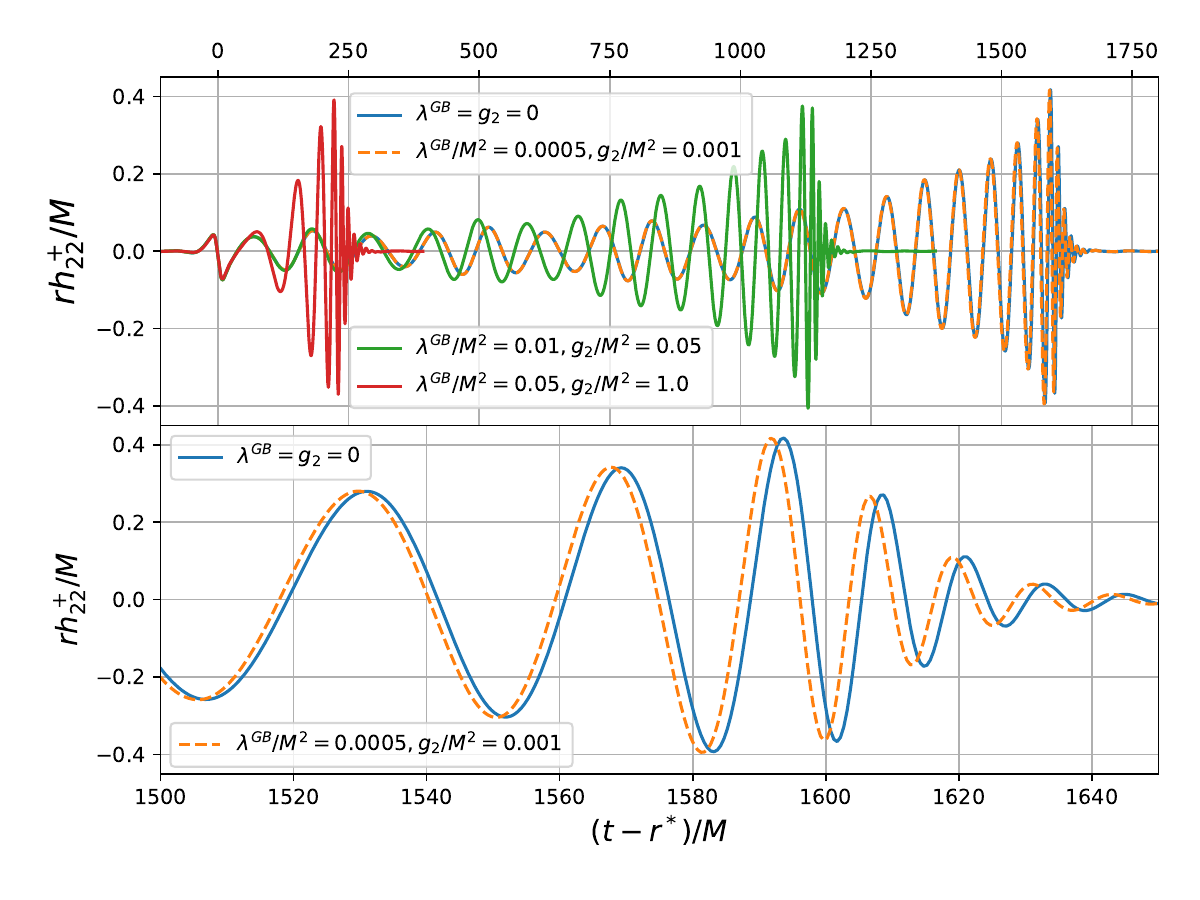}
	    \caption{Modified gravity waveforms in 4$\partial$ST with a shift-symmetric coupling. Taken from \cite{AresteSalo:2022hua}.}
        \label{fig:waveform}
    \end{figure}  
    
    \item In the work \cite{AresteSalo:2023mmd}, the studies were extended to binary mergers in theories with spin-induced scalarisation. The clouds formed are dumbbell-like in shape, as illustrated in Fig. \ref{fig:spin}.

    \begin{figure}[h!]
	    \includegraphics[width=0.99\textwidth]{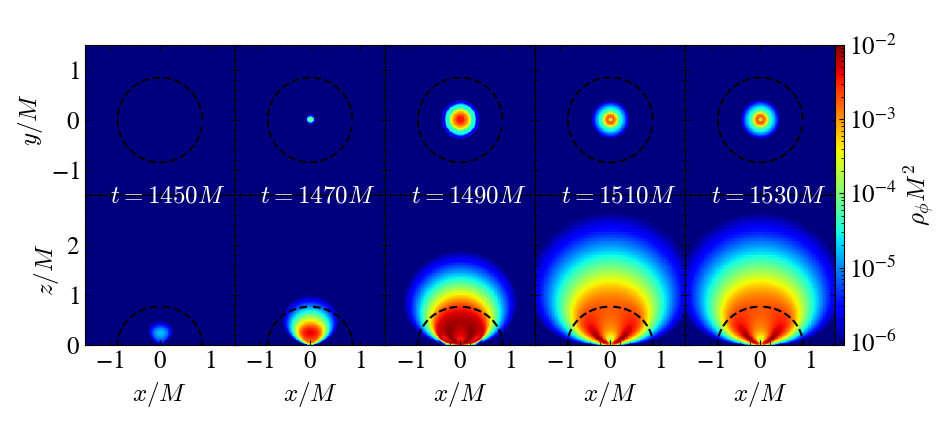}
	    \caption{The time evolution of the density of the scalar cloud that develops in Einstein-scalar-Gauss-Bonnet gravity with an exponential coupling, resulting in spin-induced scalarisation. Taken from \cite{AresteSalo:2023mmd}.}
        \label{fig:spin}
    \end{figure}  

    \item In the work \cite{Doneva:2023oww}, the dependence of the conditions for hyperbolicity and weak coupling were studied for spin-induced scalarisation, and the critical thresholds found for a number of cases, as illustrated in Fig. \ref{fig:hyperbolicity}.

    \begin{figure}[h!]
     	\includegraphics[width=0.99\textwidth]{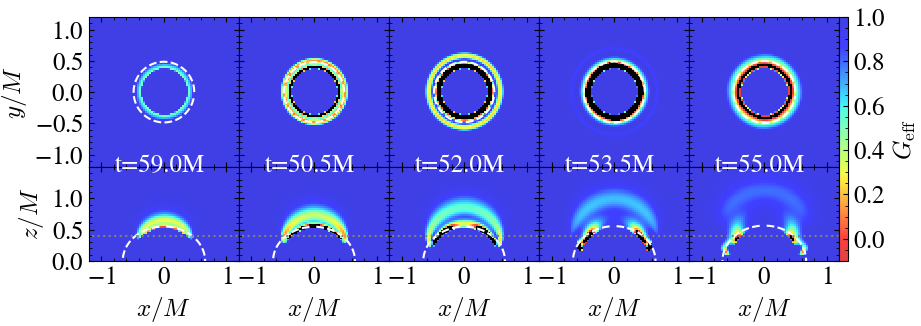}
	    \caption{The time evolution of the determinant of the effective metric in a case of spin-induced scalarisation. When the determinant is negative (in black) outside the apparent horizon (depicted with a dashed white line), the theory has become ill-posed. Taken from \cite{Doneva:2023oww}.}
        \label{fig:hyperbolicity}
    \end{figure}  
    
\end{itemize}

\section*{Acknowledgements}

We thank the entire \texttt{GRChombo} \footnote{\texttt{www.grchombo.org}} collaboration for their support and code development work. PF and KC are supported by an STFC Research Grant ST/X000931/1 (Astronomy at Queen Mary 2023-2026). PF is supported by a Royal Society University Research Fellowship  No. URF\textbackslash R\textbackslash 201026, and No. RF\textbackslash ERE\textbackslash 210291. KC is supported by an STFC Ernest Rutherford fellowship, project reference ST/V003240/1. LAS is supported by a QMUL Ph.D. scholarship.
DD acknowledges financial support via an Emmy Noether Research Group funded by the German Research Foundation (DFG) under grant no. DO 1771/1-1. 
LR is supported by a Royal Society Renewal Grant, No. URF\textbackslash R\textbackslash 201026, and a Research Expenses Enhancement Award, No. RF\textbackslash ERE\textbackslash 210291. TE is supported by the Centre for Doctoral Training (CDT) at the University of Cambridge funded through STFC. SY acknowledges the support from China Scholarship Council.

Development of the code used in this work utilised the ARCHER2 UK National Supercomputing Service\footnote{\texttt{https://www.archer2.ac.uk}} under the EPSRC HPC project no. E775, the CSD3 cluster in Cambridge under Projects No. DP128. The Cambridge Service for Data Driven Discovery (CSD3), partially operated by the University of Cambridge Research Computing on behalf of the STFC DiRAC HPC Facility. 
The DiRAC component of CSD3 is funded by BEIS capital via STFC capital Grants No. ST/P002307/1 and No. ST/ R002452/1 and STFC operations Grant No. ST/R00689X/1. DiRAC is part of the National e-Infrastructure\footnote{\texttt{www.dirac.ac.uk}}. 
Calculations were also performed using the Sulis Tier 2 HPC platform hosted by the Scientific Computing Research Technology Platform at the University of Warwick. Sulis is funded by EPSRC Grant EP/T022108/1 and the HPC Midlands+ consortium. This research has also utilised Queen Mary's Apocrita HPC facility, supported by QMUL Research-IT.
This study is in part financed by the European Union-NextGenerationEU, through the National Recovery and Resilience Plan of the Republic of Bulgaria, project No. BG-RRP-2.004-0008-C01. We acknowledge Discoverer PetaSC and EuroHPC JU for awarding this project access to Discoverer supercomputer resources. 

\bibliographystyle{h-physrev.bst}
\bibliography{biblio.bib}

\begin{thebibliography}{10}

\bibitem{LISA:2022kgy}
LISA, K.~G. Arun {\em et~al.},
\newblock Living Rev. Rel. {\bf 25}, 4 (2022), 2205.01597.

\bibitem{Perkins:2020tra}
S.~E. Perkins, N.~Yunes, and E.~Berti,
\newblock Phys. Rev. D {\bf 103}, 044024 (2021), 2010.09010.

\bibitem{Barausse:2020rsu}
E.~Barausse {\em et~al.},
\newblock Gen. Rel. Grav. {\bf 52}, 81 (2020), 2001.09793.

\bibitem{Gnocchi:2019jzp}
G.~Gnocchi, A.~Maselli, T.~Abdelsalhin, N.~Giacobbo, and M.~Mapelli,
\newblock Phys. Rev. D {\bf 100}, 064024 (2019), 1905.13460.

\bibitem{Barack:2018yly}
L.~Barack {\em et~al.},
\newblock Class. Quant. Grav. {\bf 36}, 143001 (2019), 1806.05195.

\bibitem{Baker:2014zba}
T.~Baker, D.~Psaltis, and C.~Skordis,
\newblock Astrophys. J. {\bf 802}, 63 (2015), 1412.3455.

\bibitem{Maggio:2022hre}
E.~Maggio, H.~O. Silva, A.~Buonanno, and A.~Ghosh,
\newblock (2022), 2212.09655.

\bibitem{Krishnendu:2021fga}
N.~V. Krishnendu and F.~Ohme,
\newblock Universe {\bf 7}, 497 (2021), 2201.05418.

\bibitem{LIGOScientific:2021sio}
LIGO Scientific, VIRGO, KAGRA, R.~Abbott {\em et~al.},
\newblock (2021), 2112.06861.

\bibitem{Carson:2019kkh}
Z.~Carson and K.~Yagi,
\newblock Phys. Rev. D {\bf 101}, 044047 (2020), 1911.05258.

\bibitem{Cornish:2011ys}
N.~Cornish, L.~Sampson, N.~Yunes, and F.~Pretorius,
\newblock Phys. Rev. D {\bf 84}, 062003 (2011), 1105.2088.

\bibitem{Okounkova:2022grv}
M.~Okounkova, M.~Isi, K.~Chatziioannou, and W.~M. Farr,
\newblock Phys. Rev. D {\bf 107}, 024046 (2023), 2208.02805.

\bibitem{Johnson-McDaniel:2021yge}
N.~K. Johnson-McDaniel {\em et~al.},
\newblock Phys. Rev. D {\bf 105}, 044020 (2022), 2109.06988.

\bibitem{Shiralilou:2021mfl}
B.~Shiralilou, T.~Hinderer, S.~M. Nissanke, N.~Ortiz, and H.~Witek,
\newblock Class. Quant. Grav. {\bf 39}, 035002 (2022), 2105.13972.

\bibitem{Perkins:2021mhb}
S.~E. Perkins, R.~Nair, H.~O. Silva, and N.~Yunes,
\newblock Phys. Rev. D {\bf 104}, 024060 (2021), 2104.11189.

\bibitem{Carson:2020ter}
Z.~Carson and K.~Yagi,
\newblock Phys. Rev. D {\bf 101}, 104030 (2020), 2003.00286.

\bibitem{Carson:2020cqb}
Z.~Carson and K.~Yagi,
\newblock Class. Quant. Grav. {\bf 37}, 215007 (2020), 2002.08559.

\bibitem{Doneva:2022ewd}
D.~D. Doneva, F.~M. Ramazano\u{g}lu, H.~O. Silva, T.~P. Sotiriou, and S.~S.
  Yazadjiev,
\newblock (2022), 2211.01766.

\bibitem{Horndeski:1974wa}
G.~W. Horndeski,
\newblock Int. J. Theor. Phys. {\bf 10}, 363 (1974).

\bibitem{Figueras:2020dzx}
P.~Figueras and T.~Fran\c{c}a,
\newblock Class. Quant. Grav. {\bf 37}, 225009 (2020), 2006.09414.

\bibitem{Figueras:2021abd}
P.~Figueras and T.~Fran\c{c}a,
\newblock Phys. Rev. D {\bf 105}, 124004 (2022), 2112.15529.

\bibitem{Kovacs:2020pns}
A.~D. Kov\'acs and H.~S. Reall,
\newblock Phys. Rev. Lett. {\bf 124}, 221101 (2020), 2003.04327.

\bibitem{Kovacs:2020ywu}
A.~D. Kov\'acs and H.~S. Reall,
\newblock Phys. Rev. D {\bf 101}, 124003 (2020), 2003.08398.

\bibitem{East:2020hgw}
W.~E. East and J.~L. Ripley,
\newblock Phys. Rev. D {\bf 103}, 044040 (2021), 2011.03547.

\bibitem{East:2021bqk}
W.~E. East and J.~L. Ripley,
\newblock Phys. Rev. Lett. {\bf 127}, 101102 (2021), 2105.08571.

\bibitem{East:2022rqi}
W.~E. East and F.~Pretorius,
\newblock Phys. Rev. D {\bf 106}, 104055 (2022), 2208.09488.

\bibitem{Corman:2022xqg}
M.~Corman, J.~L. Ripley, and W.~E. East,
\newblock Phys. Rev. D {\bf 107}, 024014 (2023), 2210.09235.

\bibitem{AresteSalo:2022hua}
L.~Arest\'e~Sal\'o, K.~Clough, and P.~Figueras,
\newblock Phys. Rev. Lett. {\bf 129}, 261104 (2022), 2208.14470.

\bibitem{AresteSalo:2023mmd}
L.~Arest\'e~Sal\'o, K.~Clough, and P.~Figueras,
\newblock (2023), 2306.14966.

\bibitem{Doneva:2023oww}
D.~D. Doneva, L.~Arest\'e~Sal\'o, K.~Clough, P.~Figueras, and S.~S. Yazadjiev,
\newblock (2023), 2307.06474.

\bibitem{Clough:2015sqa}
K.~Clough {\em et~al.},
\newblock Class. Quant. Grav. {\bf 32}, 245011 (2015), 1503.03436.

\bibitem{Andrade:2020dgc}
T.~Andrade, P.~Figueras, and U.~Sperhake,
\newblock JHEP {\bf 03}, 111 (2022), 2011.03049.

\bibitem{Adams:2015kgr}
M.~Adams {\em et~al.},
\newblock (2015).

\bibitem{Aurrekoetxea:2022mpw}
J.~C. Aurrekoetxea, K.~Clough, and E.~A. Lim,
\newblock Class. Quant. Grav. {\bf 40}, 075003 (2023), 2207.03125.

\bibitem{Brady:2023dgu}
S.~E. Brady, L.~Arest\'e~Sal\'o, K.~Clough, P.~Figueras, and A.~P. S,
\newblock (2023), 2308.16791.

\bibitem{Berger:1984zza}
M.~J. Berger and J.~Oliger,
\newblock J. Comput. Phys. {\bf 53}, 484 (1984).

\bibitem{Berger:1991}
M.~Berger and I.~Rigoutsos,
\newblock IEEE Transactions on Systems, Man, and Cybernetics {\bf 21}, 1278
  (1991).

\bibitem{East:2011aa}
W.~E. East, F.~Pretorius, and B.~C. Stephens,
\newblock Phys. Rev. D {\bf 85}, 124010 (2012), 1112.3094.

\bibitem{Neilsen:2007ua}
D.~Neilsen, E.~W. Hirschmann, M.~Anderson, and S.~L. Liebling,
\newblock {Adaptive Mesh Refinement and Relativistic MHD},
\newblock in {\em {11th Marcel Grossmann Meeting on General Relativity}}, pp.
  1579--1581, 2007, gr-qc/0702035.

\bibitem{Nakamura:1987zz}
T.~Nakamura, K.~Oohara, and Y.~Kojima,
\newblock Prog. Theor. Phys. Suppl. {\bf 90}, 1 (1987).

\bibitem{Shibata:1995we}
M.~Shibata and T.~Nakamura,
\newblock Phys. Rev. D {\bf 52}, 5428 (1995).

\bibitem{Baumgarte:1998te}
T.~W. Baumgarte and S.~L. Shapiro,
\newblock Phys. Rev. D {\bf 59}, 024007 (1998), gr-qc/9810065.

\bibitem{Bona:2003fj}
C.~Bona, T.~Ledvinka, C.~Palenzuela, and M.~Zacek,
\newblock Phys. Rev. D {\bf 67}, 104005 (2003), gr-qc/0302083.

\bibitem{Bernuzzi:2009ex}
S.~Bernuzzi and D.~Hilditch,
\newblock Phys. Rev. D {\bf 81}, 084003 (2010), 0912.2920.

\bibitem{Alic:2011gg}
D.~Alic, C.~Bona-Casas, C.~Bona, L.~Rezzolla, and C.~Palenzuela,
\newblock Phys. Rev. D {\bf 85}, 064040 (2012), 1106.2254.

\bibitem{Alic:2013xsa}
D.~Alic, W.~Kastaun, and L.~Rezzolla,
\newblock Phys. Rev. D {\bf 88}, 064049 (2013), 1307.7391.

\bibitem{Campanelli:2005dd}
M.~Campanelli, C.~O. Lousto, P.~Marronetti, and Y.~Zlochower,
\newblock Phys. Rev. Lett. {\bf 96}, 111101 (2006), gr-qc/0511048.

\bibitem{Baker:2005vv}
J.~G. Baker, J.~Centrella, D.-I. Choi, M.~Koppitz, and J.~van Meter,
\newblock Phys. Rev. Lett. {\bf 96}, 111102 (2006), gr-qc/0511103.

\bibitem{Evstafyeva:2022rve}
T.~Evstafyeva, M.~Agathos, and J.~L. Ripley,
\newblock Phys. Rev. D {\bf 107}, 124010 (2023), 2212.11359.

\bibitem{Witek:2018dmd}
H.~Witek, L.~Gualtieri, P.~Pani, and T.~P. Sotiriou,
\newblock Phys. Rev. D {\bf 99}, 064035 (2019), 1810.05177.

\bibitem{Loffler:2011ay}
F.~Loffler {\em et~al.},
\newblock Class. Quant. Grav. {\bf 29}, 115001 (2012), 1111.3344.

\bibitem{Schnetter:2003rb}
E.~Schnetter, S.~H. Hawley, and I.~Hawke,
\newblock Class. Quant. Grav. {\bf 21}, 1465 (2004), gr-qc/0310042.

\bibitem{Husa:2004ip}
S.~Husa, I.~Hinder, and C.~Lechner,
\newblock Comput. Phys. Commun. {\bf 174}, 983 (2006), gr-qc/0404023.

\bibitem{Richards:2023xsr}
C.~Richards, A.~Dima, and H.~Witek,
\newblock (2023), 2305.07704.

\bibitem{Elley:2022ept}
M.~Elley, H.~O. Silva, H.~Witek, and N.~Yunes,
\newblock Phys. Rev. D {\bf 106}, 044018 (2022), 2205.06240.

\bibitem{R:2022tqa}
A.~H.~K. R., E.~R. Most, J.~Noronha, H.~Witek, and N.~Yunes,
\newblock Phys. Rev. D {\bf 107}, 104047 (2023), 2212.02039.

\bibitem{Silva:2020omi}
H.~O. Silva, H.~Witek, M.~Elley, and N.~Yunes,
\newblock Phys. Rev. Lett. {\bf 127}, 031101 (2021), 2012.10436.

\bibitem{Pfeiffer:2002wt}
H.~P. Pfeiffer, L.~E. Kidder, M.~A. Scheel, and S.~A. Teukolsky,
\newblock Comput. Phys. Commun. {\bf 152}, 253 (2003), gr-qc/0202096.

\bibitem{Okounkova:2020rqw}
M.~Okounkova,
\newblock Phys. Rev. D {\bf 102}, 084046 (2020), 2001.03571.

\bibitem{Kuan:2023trn}
H.-J. Kuan {\em et~al.},
\newblock (2023), 2302.11596.

\bibitem{Yamamoto:2008js}
T.~Yamamoto, M.~Shibata, and K.~Taniguchi,
\newblock Phys. Rev. D {\bf 78}, 064054 (2008), 0806.4007.

\bibitem{Kiuchi:2017pte}
K.~Kiuchi {\em et~al.},
\newblock Phys. Rev. D {\bf 96}, 084060 (2017), 1708.08926.

\bibitem{Doneva:2022byd}
D.~D. Doneva, A.~Va\~n\'o Vi\~nuales, and S.~S. Yazadjiev,
\newblock Phys. Rev. D {\bf 106}, L061502 (2022), 2204.05333.

\bibitem{Okounkova:2019zjf}
M.~Okounkova, L.~C. Stein, J.~Moxon, M.~A. Scheel, and S.~A. Teukolsky,
\newblock Phys. Rev. D {\bf 101}, 104016 (2020), 1911.02588.

\bibitem{Okounkova:2019dfo}
M.~Okounkova, L.~C. Stein, M.~A. Scheel, and S.~A. Teukolsky,
\newblock Phys. Rev. D {\bf 100}, 104026 (2019), 1906.08789.

\bibitem{Ruchlin:2017com}
I.~Ruchlin, Z.~B. Etienne, and T.~W. Baumgarte,
\newblock Phys. Rev. D {\bf 97}, 064036 (2018), 1712.07658.

\bibitem{R:2022hlf}
A.~H.~K. R, J.~L. Ripley, and N.~Yunes,
\newblock Phys. Rev. D {\bf 107}, 044044 (2023), 2211.08477.

\bibitem{Ripley:2020vpk}
J.~L. Ripley and F.~Pretorius,
\newblock Class. Quant. Grav. {\bf 37}, 155003 (2020), 2005.05417.

\bibitem{Ripley:2019irj}
J.~L. Ripley and F.~Pretorius,
\newblock Class. Quant. Grav. {\bf 36}, 134001 (2019), 1903.07543.

\bibitem{Ripley:2019hxt}
J.~L. Ripley and F.~Pretorius,
\newblock Phys. Rev. D {\bf 99}, 084014 (2019), 1902.01468.

\bibitem{Ripley:2019aqj}
J.~L. Ripley and F.~Pretorius,
\newblock Phys. Rev. D {\bf 101}, 044015 (2020), 1911.11027.

\bibitem{OConnor:2009iuz}
E.~O'Connor and C.~D. Ott,
\newblock Class. Quant. Grav. {\bf 27}, 114103 (2010), 0912.2393.

\bibitem{Gerosa:2016fri}
D.~Gerosa, U.~Sperhake, and C.~D. Ott,
\newblock Class. Quant. Grav. {\bf 33}, 135002 (2016), 1602.06952.

\bibitem{Kuan:2021lol}
H.-J. Kuan, D.~D. Doneva, and S.~S. Yazadjiev,
\newblock Phys. Rev. Lett. {\bf 127}, 161103 (2021), 2103.11999.

\bibitem{Corelli:2022pio}
F.~Corelli, M.~De~Amicis, T.~Ikeda, and P.~Pani,
\newblock Phys. Rev. Lett. {\bf 130}, 091501 (2023), 2205.13006.

\bibitem{Corelli:2022phw}
F.~Corelli, M.~De~Amicis, T.~Ikeda, and P.~Pani,
\newblock Phys. Rev. D {\bf 107}, 044061 (2023), 2205.13007.

\end{thebibliography}

\end{document}